\documentclass[apj]{emulateapj}
\usepackage{mathptmx}

\def\gtorder{\mathrel{\raise.3ex\hbox{$>$}\mkern-14mu
             \lower0.6ex\hbox{$\sim$}}}
\def\ltorder{\mathrel{\raise.3ex\hbox{$<$}\mkern-14mu
             \lower0.6ex\hbox{$\sim$}}}

\slugcomment{Draft of \today}

\shorttitle{SGRs in nearby galaxies}

\shortauthors{Ofek}

\begin{document}

\title{SGRs in nearby galaxies: rate, luminosity function and fraction among short GRBs}

\author{Eran O. Ofek\altaffilmark{1}
}
\altaffiltext{1}{Division of Physics, Mathematics and Astronomy, California Institute of Technology, Pasadena, California 91125, USA}

\begin{abstract}

It was suggested that some of the short-duration
Gamma-Ray Bursts (GRB)
are giant flares of
Soft Gamma-ray Repeaters (SGR)
in nearby galaxies.
To test this hypothesis,
I have constructed a sample of $47$ short GRBs,
detected by the Inter-Planetary Network (IPN),
for which the position is 
constrained by at least one annulus on the
celestial sphere.
For each burst, I have checked whether its IPN $3$-$\sigma$ error region
coincides with the apparent disk of
one of $316$ bright, star-forming galaxies found within $20$~Mpc.
I find a single match of GRB~000420B with M74,
which could, however, be due to a chance coincidence.
I estimate the IPN efficiency as a function of fluence
and derive the galaxy sample completeness.
I find that
assuming there is a cut-off in the 
observed energy distribution of SGR flares at $\leq 10^{47}$~erg,
the fraction of SGRs among short GRBs
with fluence above $\sim10^{-5}$~erg~cm$^{-2}$
is $<16\%$ ($<27\%$) at the $95\%$ ($99.73\%$) confidence level.
I estimate the number of active SGRs in
each one of the galaxies in the sample,
and combine it with the distances to these galaxies,
the IPN efficiency,
and the SGR flare energy distribution (Cheng et al.),
to derive the rate of giant flares with energy above $4\times10^{46}$~erg.
I find that the rate of such giant flares is about
$(0.4-5)\times10^{-4}$~yr$^{-1}$ per SGR.
This rate is marginally consistent with the observed Galactic rate,
of a single giant flare with energy above $4\times10^{46}$~erg
in 30 years.
Comparison of the Galactic rate with 
the inferred extragalactic rate implies a gradual
cut-off (or steepening) of
the flare energy distribution at 
$\lesssim3\times10^{46}$~erg
($95\%$ confidence).
Using the Galactic SGR flare rate,
I set a lower limit of one percent on the fraction
of SGR flares among short GRBs.

\end{abstract}


\section{Introduction}
\label{Introduction}

The rate of Soft Gamma-ray Repeater (SGR) giant flares and their
fraction among short-duration Gamma Ray Bursts
(GRB; Kouveliotou et al. 1993)
are important ingredients for
the understanding of giant flares
in the context of the magnetar model 
(Duncan \& Thompson 1992; Paczynski 1992;
for a recent review see Woods \& Thompson 2006).

The large  energy release from the 2004 December 27 giant flare,
combined with some similarities between
the temporal and spectral properties
of giant flares and short GRBs, has
re-ignited the idea that some or all
of the short GRBs
are in fact SGR giant flares in nearby galaxies
(e.g., Dar 2005; Hurley et al. 2005; Palmer et al. 2005; Nakar et al. 2006).
Additional support to this hypothesis
came from the observed Galactic rate
of giant flares with energy above
$4\times10^{46}$~erg,
of one in 30~years among four active Galactic SGRs.

A limit on the fraction of SGRs among short GRBs
was obtained by Nakar et al. (2006),
who searched the error quadrilateral of six,
well-localized, Inter-Planetary Network (IPN) GRBs
for relatively nearby galaxies.
They put a upper limit of $<40\%$ ($95\%$ confidence limit; CL)
on the fraction of SGRs among bright short GRBs.

Lazzati, Ghirlanda, \& Ghisellini (2005),
inspected the spectra of 76 
Burst And Transient Source Experiment (BATSE) short bursts.
They found three GRBs
whose spectra are well described by a black body
model, and $15$ bursts for which a fit with a black body model
is not excluded.
Lazzati et al. (2005) then argued,
based on the gamma-ray light curves of these short GRBs,
that none of them were likely SGR bursts.
They derived a limit on the fraction of SGR bursts
among bright BATSE events of $<4\%$ ($95\%$ CL).
I note that the Lazzati et al. (2005) results are
based on the assumption that all the giant
flares have similar light curves.
However, this assumption is based on a small
number of observed giant flares.
If I relax the Lazzati et al. (2005)
assumption about the temporal characteristics of
giant flares,
but keep their assumption regarding the spectral
properties of SGRs (and there are up to 18 SGR
flares in their sample), their result poses an upper limit
on the fraction of SGRs among bright short GRBs
of $<35\%$, at the $95\%$ CL.

Recently, two possible detections
of extragalactic SGRs were claimed:
Crider (2006) detected a $13.8$~s periodicity in the
tail of the gamma-ray light curve of GRB~970110.
He suggested that this burst may be an SGR flare in NGC~6946.
Golenetskii et al. (2005) reported the detection of
GRB~051103, which includes the nearby galaxy M81
in its error quadrilateral
(see Ofek et al. 2006 and Frederiks et al. 2006
for follow-up observations and discussion).

Another quantity related to the fraction of SGRs among short GRBs
is the rate of SGR giant flares.
This rate
is the consequence of
the total energy available for SGRs,
which in the context of the magnetar model
is provided by the magnetic field of neutron star.

Recently, Stella et al. (2005)
argued that the rate of SGR giant flares,
with energy above $5\times10^{46}$~erg,
is about $10^{-2}$~yr$^{-1}$.
They obtained this result 
by using a uniform
prior (see however, Duncan 2001) on the observed rate.
They further argued
that about 70 giant flares,
with energy above $5\times10^{46}$~erg,
are expected during the lifetime of an SGR, and 
concluded that magnetic fields of $B\gtorder10^{16}$~G
are needed in order to explain the energy source
of SGRs.

Popov \& Stern (2005) attempted to 
find extragalactic SGRs
by searching the BATSE
catalog for short GRBs 
that spatially coincide with the Virgo cluster.
They used this search to put an 
upper limit on the rate of giant flares
with energy above $\sim10^{46}$~erg,
of $10^{-3}$~yr$^{-1}$.
However, this limit was obtained by assuming
that $T_{50}$ of all SGR giant flares is in the range
of 0.05 to 0.7~s, and by ignoring the BATSE positional uncertainty.
A similar limit was obtained by Palmer et al. (2005).

In this paper I match the IPN
(\S\ref{IPN}) short GRBs 
with nearby galaxies (\S\ref{Sample}).
Contrary to previous efforts,
I do not introduce any assumptions about the
temporal or spectral properties
of SGR giant flares.
I present, in \S\ref{Search}, the search for IPN short-duration GRBs
that spatially coincide with nearby galaxies.
In \S\ref{Rate},
I use the results of this search to calculate
the rate of SGR giant flares and confront it with
the rate estimated based on the Galactic SGRs.
I finally discuss the results in \S\ref{Disc}.

\section{The IPN sensitivity}
\label{IPN}

The Inter-Planetary Network (IPN; 
e.g., Laros et al. 1997; Hurley et al. 1999)
is a set of gamma-ray detectors on board several spacecraft
in the solar system.
The IPN uses GRB photon arrival times
to triangulate their celestial positions.
Each pair of spacecraft constrain the GRB position to
an annulus on the celestial sphere.
The area of IPN $3$-$\sigma$ error regions
is typically two orders of magnitude
smaller than BATSE error circles.
In fact, IPN positions were used to calibrate
BATSE positional errors (Briggs et al. 1999).

\begin{figure}
\centerline{\includegraphics[width=8.5cm]{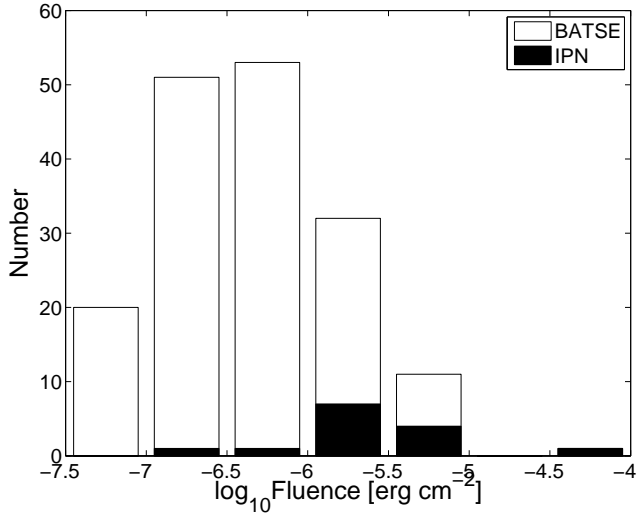}}
\caption {Histogram of BATSE GRBs integrated four-channel
fluence,
for short-duration bursts triggered by BATSE (empty bars),
and BATSE-triggered short-duration GRBs that were detected by at least two IPN
detectors and thus constrained by at least one annulus (filled bars).
\label{BATSE_IPN_Fluence_Hist} }
\end{figure}

Here I have use the IPN catalog\footnote{http://www.ssl.berkeley.edu/ipn3/interpla.html; version 2005 December 17 -- For reference, this IPN catalog version is available from htpp://astro.caltech.edu/$\sim$eran/GRB/IPN/NearbyGal/CatIPN.txt.ver17122005}
containing all IPN-triangulated GRBs observed
from 1990 November 12, to 2005 October 31\footnote{This catalog does not contain GRB~051103 which may have originated in the nearby galaxy M81
(Ofek et al. 2006; Frederiks et al. 2006).}.
Figure~\ref{BATSE_IPN_Fluence_Hist} shows a
histogram of the integrated four-channel fluence
for BATSE-triggered short-duration GRBs ($T_{90}<2$~s; empty bars),
as well as all short-duration GRBs triggered both by BATSE
and an additional IPN
detector (i.e., that have at least one annulus constraint; filled bars).

\begin{figure}
\centerline{\includegraphics[width=8.5cm]{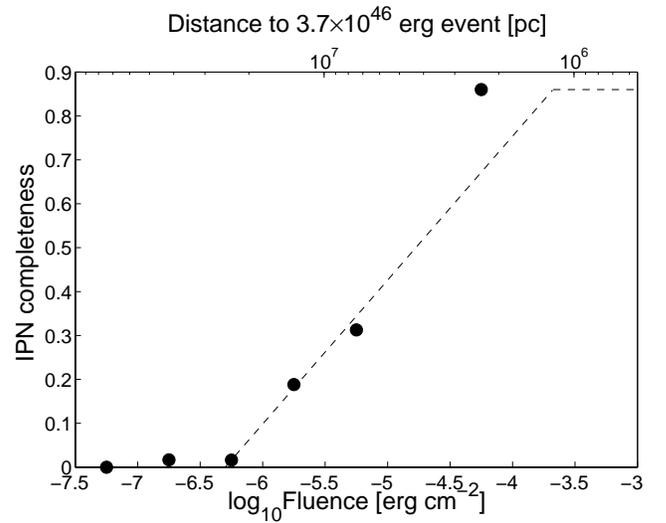}}
\caption {The all-sky IPN approximate completeness
as calculated from the ratio of the two histograms in Fig.~\ref{BATSE_IPN_Fluence_Hist},
multiplied by the exposure completeness ($86\%$; see text).
The upper x-axis shows the corresponding distance to a $3.7\times10^{46}$~erg
event, which is the energy of the 2004 December 27 flare
assuming a distance of $15$~kpc to SGR~1806$-$20 (Hurley et al. 2005).
The dashed line is an approximation
to this completeness function (Eq.~\ref{FunC}).
\label{IPN_completness} }
\end{figure}

Figure~\ref{IPN_completness} shows the approximate
all-sky IPN completeness
as a function of BATSE integrated four-channel fluence.
Completeness here is calculated from the ratio of the
two histograms in Fig.~\ref{BATSE_IPN_Fluence_Hist},
multiplied by the exposure completeness
of the Konus/Wind and BATSE short-duration GRB sample
that is presented in the next section.
I estimated the exposure completeness to be about
$86\%$ ($=[4\times0.48 + 11\times1.0]/15$).
This is based on the fact
Konus/Wind started to work only in 1994
(11~yrs out of 15~yrs), and the
exposure completeness of BATSE,
which worked 4~yrs without Konus/Wind,
is about $48\%$\footnote{Calculated from the BATSE all-sky exposure
map http://cossc.gsfc.nasa.gov/docs/cgro/cossc/batse/4Bcatalog/4b\_exposure.html}.
The dashed line shows an approximation to this completeness function
given by:
\begin{equation}
C(F) = \left\{ \begin{array}{ll}
         0,                          &  F\leq5\times10^{-7}               \\
         0.328\log_{10}{F}+2.065,    &  5\times10^{-7}<F<2.1\times10^{-4} \\
         0.86,                       &  F\geq2.1\times10^{-4},                          
     \end{array} \right.
\label{FunC}
\end{equation}
where $F$ is the fluence.
BATSE sensitivity is almost complete
down to a fluence of about $10^{-6}$~erg~cm$^{-2}$,
in which the IPN efficiency drops to about ten percent.
Therefore, the sensitivity
function (Eq.~\ref{FunC}) is an approximation to the IPN
absolute mean efficiency.
However, I will show in \S\ref{Sample} that
the results are not very sensitive to the exact shape of this
completeness function.

\section{Selection of the GRB and galaxy samples}
\label{Sample}


I selected from the IPN catalog all GRBs 
with a $3$-$\sigma$ error region
constrained by at least one
annulus with semi-width smaller than one~degree (1260).
I further selected from this list all GRBs that were triggered by
BATSE (752) and
have $T_{90}<2.0$~s (29 events).
The $T_{90}$ duration at which a GRB has an equal
probability of being a short- or long-duration GRB is
about $T_{90}=5$~s (Donaghy et al. 2006).
As SGR giant flares have durations shorter than 2~s,
the upper limit on the fraction of SGR giant flares among short-duration
GRBs derived in this paper is conservative.

\begin{figure}
\centerline{\includegraphics[width=8.5cm]{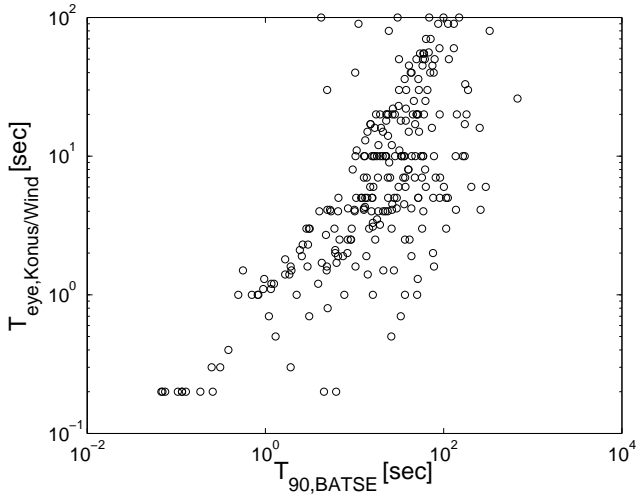}}
\caption {The BATSE $T_{90}$ vs. the estimated duration
(by eye from Konus/Wind light curve), $T_{eye}$, of the GRBs.
\label{T90_Teye} }
\end{figure}

In addition, I searched for short-duration GRBs among
the Konus/Wind satellite observations (Cline et al. 2003).
A catalog of short GRBs detected by Konus/Wind was published
by Mazets et al. (2004), however, some of the GRBs in this
catalog were classified as long GRBs by BATSE.
Therefore, instead of using this catalog I used the following
statistical approach:
I inspected by eye the Konus/Wind light curves
for which an IPN annulus constraint is available,
and selected all the bursts that have
a total duration of less than two~seconds.
Due to the differences between the BATSE and
Konus/Wind detectors, this sample is contaminated by long
GRBs (which appear like short GRBs for the Konus/Wind detector).
Figure~\ref{T90_Teye} shows the BATSE $T_{90}$ vs. the duration
estimated by eye
($T_{eye}$) of GRBs that were triggered by both BATSE and Konus/Wind.
I found 83 Konus/Wind bursts with $T_{eye}\leq 2$~s and
annulus semi-width of less than one~deg.
Of these 83 events,
37 were detected by BATSE, of which 23 were
classified by BATSE as long-duration bursts (i.e., $T_{90}>2$~s).
Therefore, the Konus/Wind sample
has $\sim60\%$ ($23/37$) contamination by long GRBs.
Given this contamination factor,
I estimate that the actual number
of short  bursts, among these
46 remaining events, is $18_{-4}^{+5}$.
Finally, I compiled a list of 75 events:
29 BATSE events with $T_{90}<2$~s and 46 Konus/Wind events with $T_{eye}<2$~s.
%
%
This list contains about $47$ ($\sim18+29$) short bursts.


Next, I selected a sample of bright star-forming galaxies within $20$~Mpc
from a modified version of the Tully (1988)
nearby galaxies catalog.
The selected galaxies have
absolute magnitude $M_{B}\le-18$,
reside at distances less than 20~Mpc, and have
morphological types other than E or S0.
Introducing the last criterion, I assume that SGRs
form predominantly in young stellar populations
(see however, Levan et al. 2006).
I excluded from this list the Large Magellanic Cloud (LMC).
The nearby galaxy sample contains 316 galaxies,
which are listed along with their basic properties
in Table~\ref{Table-Dist}.
Also listed in this table are the
far Infra-Red (IR) fluxes and estimated supernova (SN) rates in
these galaxies derived in \S\ref{Rate}.
\begin{deluxetable*}{lllrccccc}
\tablecolumns{9}
\tabletypesize{\scriptsize}
\tablewidth{0pt}
\tablecaption{Nearby bright galaxies}
\tablehead{
\colhead{Name} &
\colhead{R.A. J2000.0 Dec.} &
\colhead{$B_{T}$\tablenotemark{a}} &
\colhead{$M_{B}$\tablenotemark{b}$^{,}$\tablenotemark{d}} &
\colhead{Ang. Diam.} &
\colhead{Distance\tablenotemark{c}$^{,}$\tablenotemark{d}} &
\colhead{Type\tablenotemark{e}} &
\colhead{IR Flux\tablenotemark{f}} &
\colhead{SN rate\tablenotemark{g}} \\
\colhead{} &
\colhead{} &
\colhead{[mag]} &
\colhead{[mag]} &
\colhead{[arcmin.]} &
\colhead{[Mpc]} &
\colhead{} &
\colhead{[Jy]} &
\colhead{[yr$^{-1}$]}
}
\startdata
N 7814  & $00:03:18$ $+16:09:00$ & $10.92$ & $-20.12$ & $   5.9$ & $ 16.2$ & $ 2$ & $     0.000$ & $0.00000$ \\
N   14  & $00:08:48$ $+15:49:00$ & $12.10$ & $-18.59$ & $   2.9$ & $ 13.7$ & $10$ & $     0.000$ & $0.00000$ \\
N   55  & $00:14:54$ $-39:11:00$ & $ 7.50$ & $-18.22$ & $  34.1$ & $  1.4$ & $ 9$ & $   202.119$ & $0.00245$ \\
N  178  & $00:39:06$ $-14:11:00$ & $12.77$ & $-18.70$ & $   2.1$ & $ 19.7$ & $ 9$ & $     0.000$ & $0.00000$ \\
N  224  & $00:42:42$ $+41:16:00$ & $ 3.56$ & $-20.82$ & $ 193.2$ & $  0.7$ & $ 3$ & $   793.507$ & $0.00279$ \\
N  247  & $00:47:06$ $-20:45:00$ & $ 8.66$ & $-18.10$ & $  22.2$ & $  2.2$ & $ 7$ & $     8.335$ & $0.00026$ \\
N  253  & $00:47:36$ $-25:18:00$ & $ 7.34$ & $-20.20$ & $  25.8$ & $  3.2$ & $ 5$ & $  3100.177$ & $0.20000$ \\
N  278  & $00:52:06$ $+47:33:00$ & $10.75$ & $-19.76$ & $   2.7$ & $ 12.6$ & $ 3$ & $   106.233$ & $0.10603$ \\
0102-06 & $01:05:06$ $-06:13:00$ & $11.70$ & $-19.15$ & $   4.4$ & $ 14.8$ & $ 7$ & $     5.152$ & $0.00703$ \\
N  406  & $01:07:24$ $-69:53:00$ & $11.90$ & $-19.44$ & $   2.7$ & $ 18.5$ & $ 3$ & $     7.853$ & $0.01685$ \\
...     &                        &         &          &          &         &      &              &           \\
N  628  & $01:36:42$ $+15:47:00$ & $ 9.61$ & $-20.47$ & $   7.0$ & $ 10.4$ & $ 5$ & $    19.736$ & $0.01331$ \\
N 7331  & $22:37:06$ $+34:26:00$ & $ 9.67$ & $-21.26$ & $   9.7$ & $ 15.3$ & $ 4$ & $   131.932$ & $0.19339$ \\
\enddata
\tablenotetext{a}{Apparent blue magnitude, $B_{T}$,corrected for Galactic extinction, and for the external galaxy extinction (see Tully 1988).}
\tablenotetext{b}{Absolute blue magnitude as calculated from the corrected distance and apparent extinction-corrected magnitude.}
\tablenotetext{c}{Distances are based on recession velocities, assuming $H_{0}=70$~km~s$^{-1}$~Mpc$^{-1}$.}
\tablenotetext{d}{The distances listed in the Tully (1988) catalog are based on galaxy recession velocities, assuming an Hubble constant
$H_{0}=75$~km~s$^{-1}$~Mpc$^{-1}$. I corrected all the distances in this catalog by a factor of ($75/70$; i.e., to convert to $H_{0}=70$~km~s$^{-1}$~Mpc$^{-1}$),
and recalculated the absolute magnitude accordingly.}
\tablenotetext{e}{Galaxy type, see Tully (1988) for details.}
\tablenotetext{f}{The total observed far IR flux, $2.58 f_{60} + f_{100}$, calculated
by summing the fluxes of the IRAS sources found within the galaxy radius from the galaxy position.}
\tablenotetext{g}{SN rate estimated by normalizing the absolute far IR flux to that of NGC~253, and assuming this galaxy has $0.2$~SN~yr$^{-1}$ (Pietsch et al. 2001).}
\tablecomments{The first ten lines of Table~\ref{Table-Dist}. For convenience, I list also M74 (NGC~626) and NGC~7331
which are mentioned in this paper. The entire table is available via the electronic version.}
\label{Table-Dist}
\end{deluxetable*}

\subsection{The galaxy sample completeness}
\label{gal_comp}

There are three factors which affect the completeness of this
galaxy sample:
(1) galaxies behind the Galactic plane;
(2) star formation in galaxies not included in the sample;
and (3) the fact that the sample is volume limited.
I estimated these incompleteness factors below.
%


To estimate the completeness of the galaxy sample (Table~\ref{Table-Dist})
due to Galactic obscuration,
I calculated the number of galaxies in this catalog as
a function of Galactic latitude, $b$.
I then extrapolated
the number of galaxies from the unobscured regions (i.e., $b\sim30$~deg)
to the obscured regions.
Based on this extrapolation,
I estimated the completeness of the galaxy sample
to be about $\epsilon_{G}=90\%$.


Next, I estimated, $\epsilon_{SFR}$, the fraction of
star formation within galaxies with $M_{B}\le-18$
relative to the total star formation within 20~Mpc.
I calculated the far Infra-Red (IR)
flux\footnote{Based on Infra-Red Astronomical Satellite (IRAS) point source catalog, version 2.0, IPAC (1986)}
of the galaxies, defined by $2.58 f_{60} + f_{100}$,
where $f_{60}$ and $f_{100}$ are the $60$~$\mu$m and $100$~$\mu$m fluxes, respectively.
This quantity is a good estimator of the star formation rate
in galaxies (Helou et al. 1988).
Given the galaxy distances, I summed the absolute far-IR flux
of all the galaxies (excluding E0/S0) brighter than $M_{B}<-18$
within ten~Mpc
and divided it by the absolute far-IR flux
of all galaxies (excluding E0/S0) brighter than $M_{B}<-12$.
Note that I use a distance cut of ten~Mpc
in order to avoid incompleteness due to missing faint galaxies.
I found that $\epsilon_{SFR}=95\%$.


The galaxy sample contains galaxies within $20$~Mpc. However,
SGR giant flares can be observed to larger distances.
To estimate the survey volume completeness,
I divide the expected number of SGR flares
(assuming constant number density of SGRs)
that can be observed by the IPN
up to a distance of $r_{max}$, by the
total observable number of IPN SGR flares:
\begin{equation}
\epsilon_{V}=\frac{\int_{0}^{r_{max}}{4\pi r^{2}}\int_{0}^{E_{max}}{C\left( \frac{E}{4\pi r^2} \right) E^{-\alpha} dE dr}}{\int_{0}^{\infty}{4\pi r^{2}}\int_{0}^{E_{max}}{C\left( \frac{E}{4\pi r^2} \right) E^{-\alpha} dE dr}},
\label{Eff_r}
\end{equation}
where $r$ is the distance, $C(F)$ is the
completeness function (Eq.~\ref{FunC}),
and $E_{max}$ is the break in the flare energy distribution.
I weighted the number of flares of energy $E$, $dN/dE$,
by the empirical $E^{-\alpha}$ factor,
with $\alpha=5/3$ (Cheng et al. 1996).

\begin{figure}
\centerline{\includegraphics[width=8.5cm]{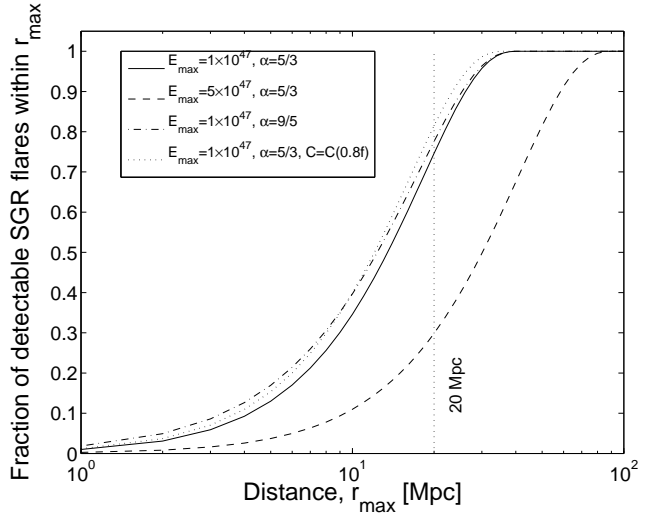}}
\caption {The survey volume completeness,
$\epsilon_{V}$ (Eq.~\ref{Eff_r}),
as a function of distance $r_{max}$, for four cases:
$E_{max}=10^{47}$~erg, $\alpha=5/3$ (solid line);
$E_{max}=5\times10^{47}$~erg, $\alpha=5/3$ (dashed line);
$E_{max}=10^{47}$~erg, $\alpha=9/5$ (dashed-dotted line);
and $E_{max}=10^{47}$~erg, $\alpha=5/3$, and a modified completeness function
(Eq.~\ref{FunC}; $C=C(0.8F)$; dotted line).
\label{IPN_completness_r} }
\end{figure}
\begin{figure}
\centerline{\includegraphics[width=8.5cm]{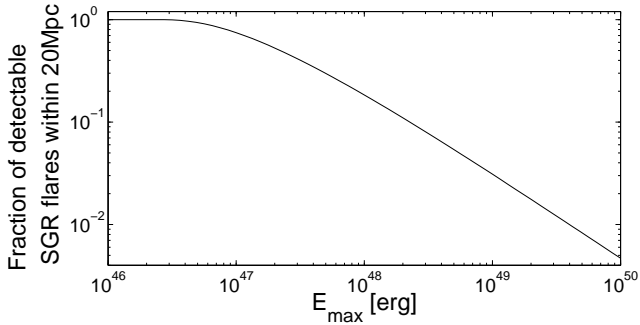}}
\caption {The survey volume completeness,
$\epsilon_{V}$ (Eq.~\ref{Eff_r}), for $r_{max}=20$~Mpc,
as a function of the break in the flare energy distribution, $E_{max}$
(assuming $\alpha=5/3$).
\label{IPN_completness_20Mpc} }
\end{figure}
Figure~\ref{IPN_completness_r} shows $\epsilon_{V}$
as a function of the distance $r_{max}$, for four
different combinations of $E_{max}$, $\alpha$ and $C(F)$
(see legend).
It is clear that the volume completeness is not sensitive to small changes in
$\alpha$, nor to the exact shape of the
completeness function.
However,
the completeness,
$\epsilon_{V}(r_{max}=20$~Mpc$)$ is sensitive to the break energy,
$E_{max}$.
Fig~\ref{IPN_completness_20Mpc} shows the volume completeness,
$\epsilon_{V}(r_{max}=20$~Mpc$)$,
as a function of $E_{max}$.
A reasonable value for $E_{max}$ is $10^{47}$~erg, which
roughly corresponds to
the total magnetic energy available for a $B=10^{15}$~G neutron star.
Assuming $E_{max}=10^{47}$~erg,
the volume-completeness of our survey is
$\epsilon_{V}(r_{max}=20$~Mpc$)=0.75$, while if
$E_{max}=5\times10^{47}$~erg, then the survey completeness
drops to $30\%$
(see however \S\ref{Rate}).

\section{The Search for SGR flares from Nearby Galaxies}
\label{Search}

For each burst in 
the short GRB sample, I searched for overlap between
its IPN $3$-$\sigma$ error region and
the apparent disk of one of the galaxies in Table~\ref{Table-Dist}.
The IPN constraints include one or more annuli
on the celestial sphere.
In addition to the annuli, the IPN database 
occasionally contains constraints from individual instruments:
ecliptic latitude constraints from Konus/Wind;
error radii from various satellites; and/or planet
blocking regions (i.e., in cases that Earth or Mars
blocked part of the detector field of view).
Some of these error regions
(e.g., BATSE; WATCH Sazonov et al. 1998; COMPTEL Kippen et al. 1998;
EGRET Gonzalez et al. 2004; PHEBUS Tkachenko et al. 2002;
SIGMA Claret et al. 1994;
BeppoSAX Guidorzi et al. 2004; HETE Vanderspek et al. 2003)
are given at the $1$-$\sigma$ confidence.
For these error radii I added known systematic errors
(i.e., $1.6$~deg for BATSE; Briggs et al. 1999) and translated 
all the errors, which I assumed are
circular\footnote{Note that error regions for GRB positions
obtained by some of the spacecraft are not circular}
and normally distributed, to $3$-$\sigma$ errors.

To avoid incompleteness due to the positional inaccuracies of the
Tully (1988) catalog,
the galaxy diameter used
in the search was increased by one arcminute.

\begin{figure}
\centerline{\includegraphics[width=8.5cm]{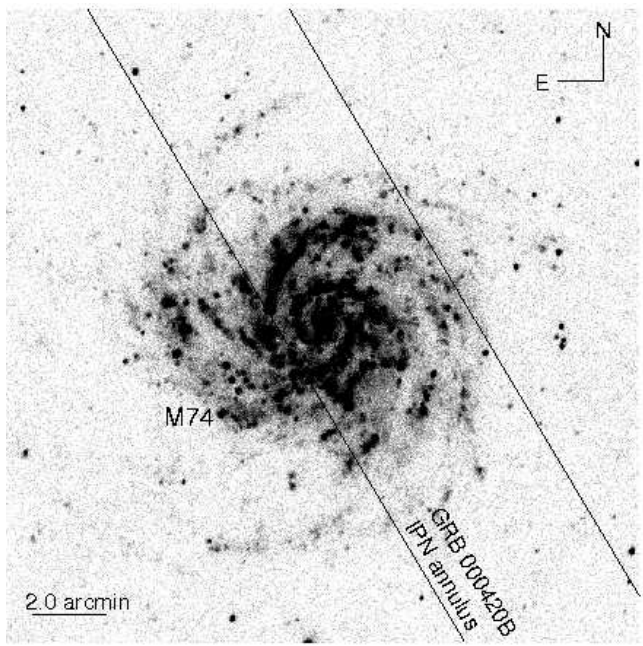}}
\caption {A GALEX near UV-band image of the Sc-type galaxy M74 (NGC~628).
The parallel lines mark the section of the IPN annulus of GRB~000420B
that passes through the galaxy.
The IPN constraints for this burst include an annulus
whose parameters are:
annulus center, $\alpha=303.76$~deg, $\delta=-29.4$~deg (J2000.0),
annulus radius of $89.63$~deg, and semi-width
of $0.04$~deg;
and a Konus/Wind ecliptic latitude constraints which indicates
that the GRB ecliptic latitude is between $+2$~deg to $+22$~deg ($3$-$\sigma$;
The ecliptic latitude of M74 is $+5.2$~deg).
The total area of this annulus segment
is $3.25$ deg$^{2}$.
\label{M74} }
\end{figure}

\begin{figure}
\centerline{\includegraphics[width=8.5cm]{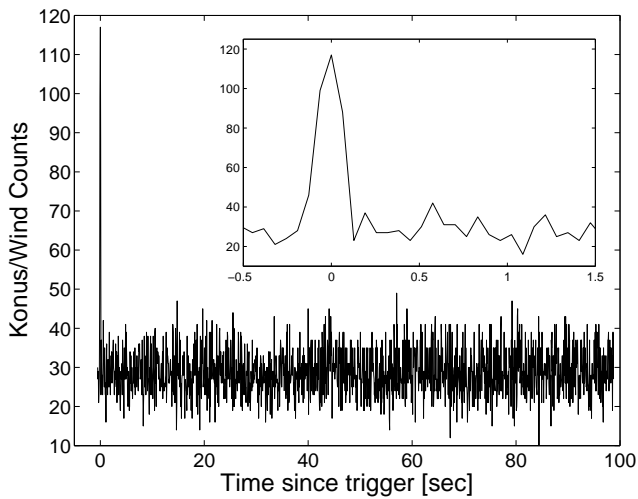}}
\caption {The Konus/Wind gamma-ray light curve of
GRB~000420B, which might have originated in M74.
The GRB shows a single, short ($0.3$~s width), peak.
\label{GRB000420_LC} }
\end{figure}

I have found two matches between the short GRB candidates
and the galaxies listed in Table~\ref{Table-Dist}.
The first, GRB~000420B, coincides with the position of
the nearby Sc-type galaxy M74.
Fig~\ref{M74} shows a Galaxy Evolution Explorer (GALEX) near UV-band image,
emphasizing the star-forming regions in M74,
with the IPN annulus overlayed (solid lines).
GRB~000420B was detected by Konus/Wind at
UTC 2000 April 20 11:44:31\footnote{On April 20th, M74 is about six~deg from the Sun}, 
and its gamma-ray light curve is shown in Fig.~\ref{GRB000420_LC}.
Note that I did not detect, using a Scargle (1982) periodogram,
any periodic signal in the light curve within $\sim100$~s
after the GRB spike.

\begin{figure}
\centerline{\includegraphics[width=8.5cm]{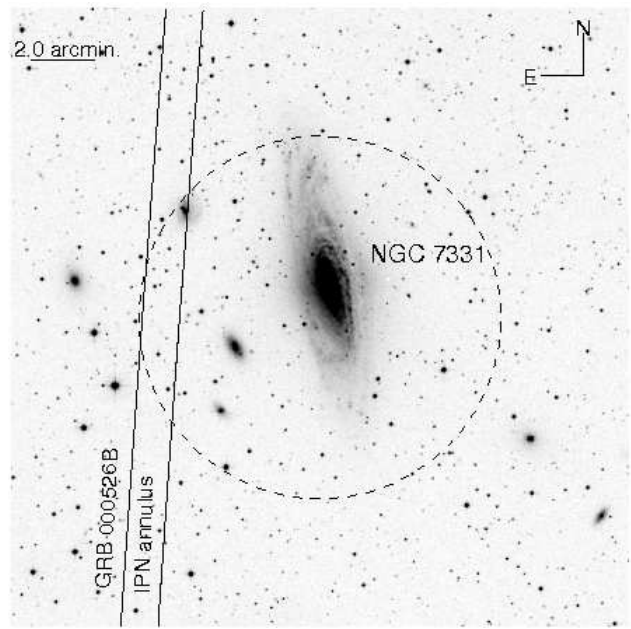}}
\caption {R-band image of the galaxy NGC~7331 from the Palomar Observatory Sky Survey II.
The parallel solid lines mark the section of the IPN annulus of GRB~000526B that passes
near this galaxy.
The center of the dashed-line circle around NGC~7331 shows the actual position of NGC~7331 in the
Tully (1988) catalog, which is accurate only to about one arcminute.
To avoid any incompleteness caused by the positional inaccuracies, I
increased the galaxy radii by one arcminute
compared to the radii listed in Tully (1988).
The circle size corresponds to the
increased-galaxy radius.
\label{NGC7331} }
\end{figure}

The second match, GRB~000526B coincides with the
position of the galaxy NGC~7331.
However, as seen in Figure~\ref{NGC7331},
this galaxy has an elongated shape
(while in the search I assumed the galaxies are round)
and
the IPN annulus does not coincide with
the galaxy apparent disk or any star forming regions
within this galaxy.
I note there is a smaller galaxy within the NGC~7331 group
that does fall within the IPN annulus.
However, this galaxy does not
pass the absolute magnitude cut, and the a-priori probability
to find an SGR flare from such a galaxy is small.
Therefore, I conclude that GRB~000526B is unlikely to be
associated with NGC~7331.

I have estimated the probability
for a chance coincidence between the short-GRB IPN
error regions and the galaxy sample.
This was done by randomizing the positions 
of the 316 galaxies over the celestial sphere,
and searching for
a match with one of the IPN short-GRBs in the list.
I repeated this simulation
1000 times and found that 
the expectancy number of matches is $\sim3$ and 
the probability to find
two matches
is $\sim15\%$, and three or more matches is $\sim76\%$.
Note that since the actual galaxy positions are correlated and
the IPN constraints are coordinate dependent
(e.g., Konus/Wind ecliptic latitude constraint)
this simulation does not mimic the search process accurately.
However, it indicates
that such a chance coincidence is possible.
Therefore, I cannot securely identify GRB~000420B with M74.

Given the product of the completeness factors,
$\epsilon_{G}\epsilon_{SFR}\epsilon_{V}\approx64\%$ (see \S\ref{Sample}), and
conservatively assuming that the search
yielded a single SGR candidate among 47 short GRBs,
I put an upper limit on the fraction of
SGRs among short GRBs
of $<16\%$ ($<27\%$),
at the one sided $95\%$ ($99.73\%$) CL (Gehrels 1986).
%
%
%
Assuming that GRB~000420B is indeed an SGR flare in M74,
the fraction of SGRs among short GRBs is about $3\%$
($=\frac{1}{47\epsilon_{SFR}\epsilon_{G}\epsilon_{V}}$).
%
%

\section{The rate of giant flares}
\label{Rate}

A closely related quantity to the fraction of SGR giant flares
among short GRBs is the rate of SGR giant flares,
which I estimate below.

Among the four active SGRs in the Galaxy (including the LMC),
a single giant flare with energy above $3.7\times10^{46}$~erg
was observed in the last 30~years.
Therefore,
the derived rate of giant flares
with energy above $3.7\times10^{46}r_{15}^{2}$~erg (Hurley et al. 2005)
is $(120_{-100}^{+5100})^{-1}$~yr$^{-1}$ per SGR
($95\%$ confidence interval; CI; Gehrels 1986),
where $r_{15}$ is the distance to SGR~1806$-$20 in 15~kpc units (Corbel \& Eikenberry 2004).
%
%

Next, using the results of \S\ref{Search},
I estimate the rate of SGR giant flares, and
compare it with the rate 
derived from the Galactic SGR giant flare.
Assuming that all the SGRs have identical
flare properties (i.e., the same cut-off energy $E_{max}$ and $\alpha$),
then the rate of giant flares,
with energy above $E_{gf}$, per year per SGR, $f_{gf}$, is given by:
\begin{equation}
f_{gf}(\ge E_{gf},E_{max})=\frac{\sum_{i}{N_{o,i}}}{\Delta{T}} \frac{ \int_{E_{gf}}^{E_{max}}{E^{-\alpha}dE} }{ \sum_{i}{ N_{i}^{sgr} \int_{0}^{E_{max}}{E^{-\alpha} C\left( \frac{E}{4\pi r_{i}^{2}}   \right)dE    }  } },
\label{Rate_SGR}
\end{equation}
where $\Delta{T}$ is the time span of observations ($15$~yrs),
$N_{o,i}$ is the observed number of SGR flares
in the $i$-th galaxy ($\sum_{i}{N_{o,i}}\leq1$),
$\alpha=5/3$ (Cheng et al. 1996),
$N_{i}^{sgr}$ is the number of active SGRs in the $i$-th galaxy,
and $C$ is the completeness function (Eq.~\ref{FunC}).
To estimate $N_{i}^{sgr}$, I use the total
far-IR flux (see \S\ref{Sample})
of the galaxies (Table~\ref{Table-Dist}) as an estimator for the star formation rate (Helou et al. 1988).
I converted the far-IR flux to the SN rate, $R_{i}^{sn}$,
by normalizing the far-IR flux of each galaxy to that of NGC~253,
which I assumed has $0.2$~SN~yr$^{-1}$ (Pietsch et al. 2001).
The far-IR flux along with the derived SN rate
for each galaxy in this sample are listed in Table~\ref{Table-Dist}.
Finally, the number of SGRs in the $i$-th galaxy, $N_{i}^{sgr}$, is:
\begin{equation}
N_{i}^{sgr} \cong N_{mw}^{sgr} \frac{R_{i}^{sn}}{R_{mw}^{sn}},
\label{N_i_sgr}
\end{equation}
where, $N_{mw}^{sgr}$ is the number of active SGRs in the Milky-Way
galaxy (four), and the SN rate in our Galaxy, $R_{mw}^{sn}$,
was set to $0.03$~yr$^{-1}$.

\begin{figure*}
\centerline{\includegraphics[width=16.2cm]{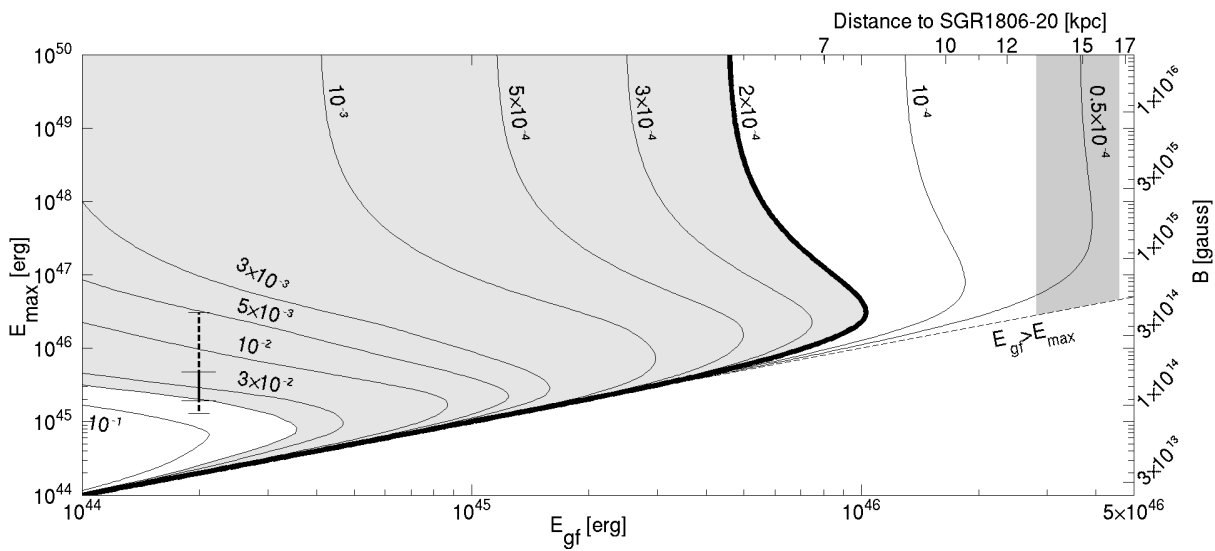}}
\caption {Contours of equal $f_{gf}/\sum_{i}{N_{o,i}}$,
as a function of the giant flare energy, $E_{gf}$, and the cut-off energy, $E_{max}$.
The labels on the contours are the SGR giant flares rate in yr$^{-1}$
(i.e., rate of events for which $E>E_{gf}$).
The triangular area in the bottom-right is the unphysical region in which $E_{gf}>E_{max}$.
The $95\%$ confidence region of
giant flare rate (per SGR), with $E>3.7\times10^{46}$~erg,
as implied by the Galactic SGRs observations,
(i.e., $0.0083_{-0.0081}^{+0.039}$~yr$^{-1}$), is shown as the light-gray region
on the left side of the Figure, and the $2$-$\sigma$ lower limit
(i.e., approximately $2\times10^{-4}$~yr$^{-1}$) is marked by a bold line.
The estimated energy of the 2004 December 27 giant flare,
$(3.7\pm0.9)\times10^{46}$~erg assuming a distance
of $15$~kpc to SGR~1806$-$20, is marked as a dark-gray region on the right side of the Figure.
The error-bar on the left-hand side marks the rate of Galactic SGR giant flares,
with $E>2\times10^{44}$~erg.
The $1$-$\sigma$~CI is marked by a solid line and
the $2$-$\sigma$~CI is marked by a dashed-line.
For convenience, I plot on the upper axis of the Figure
the distance to SGR~1806$-$20, which corresponds to the energy $E_{gf}$,
of the 2004 December 27 giant flare.
On the right-axis I plot the approximate magnetic field, $B$, corresponding
to $E_{max}$ ($=\frac{B^{2}}{8\pi} \frac{4}{3}\pi R_{ns}^{3}$),
where I set the neutron star radius, $R_{ns}$, to ten~km.
\label{E_gf_E_max} }
\end{figure*}

Now I can compare the
observed Galactic rate of giant flares with the rate
implied by the extragalactic survey.
Figure~\ref{E_gf_E_max} shows
contours of equal $f_{gf}/\sum_{i}{N_{o,i}}$
(i.e., the rate $f_{gf}$ in Eq.~\ref{Rate_SGR} calculated
assuming a single SGR flare was detected
in the sample, $\sum_{i}{N_{o,i}}=1$),
as a function of the giant flare energy $E_{gf}$ and the cut-off energy, $E_{max}$.
The labels on the contours denote the SGR giant flare rate in yr$^{-1}$ per SGR.
The $95\%$ confidence region 
on the rate of
giant flares,
with energy above $3.7\times10^{46}$~erg,
as implied by the Galactic SGR observation
(i.e., $0.0083_{-0.0081}^{+0.039}$~yr$^{-1}$ per SGR) 
is shown as the light-gray region
on the left side of the Figure, and the $2$-$\sigma$ lower limit
(i.e., approximately $2\times10^{-4}$~yr$^{-1}$) is marked as a bold line.
The estimated energy
with its uncertainty,
of the 2004 December 27 giant flare,
assuming a distance of $15$~kpc to SGR~1806$-$20,
is marked as dark-gray region on the right side of the figure.
For convenience, I plot on the upper axis of Figure~\ref{E_gf_E_max}
the distance to SGR~1806$-$20 which corresponds to the energy
of the 2004 December 27 giant flare.
Plotted on the right-axis is the approximate magnetic field, $B$,
corresponding
to $E_{max}$ ($=\frac{B^{2}}{8\pi} \frac{4}{3}\pi R_{ns}^{3}$,
where I set the neutron star radius, $R_{ns}$, to ten~km).

Apparently, there is a conflict between
the Galactic rate of SGR giant flares
and the rate implied by the extragalactic survey
which predicts
a giant flare rate of $\sim0.5\times10^{-4}$~yr$^{-1}$
(assuming $E_{gf}\sim3.7\times10^{46}$~erg).
There are several ways to explain this contradiction:
The observed number of SGR flares in the search, $\sum_{i}{N_{o,i}}=1$, 
is subject to Poisson error $1_{-0.98}^{+4.7}$ ($95\%$ CL);
The estimate of $N_{i}^{sgr}$ (Eq.~\ref{N_i_sgr}) is subject to errors through
the SN rate estimate,
and to Poisson error in the number of active SGRs in
the Galaxy ($4_{-3}^{+6}$; $95\%$ CL);
The distance to SGR~1806$-$20 may be smaller than $15$~kpc
(e.g., 12kpc, Figer et al. 2004; 6--10~kpc, Cameron et al. 2005),
and therefore shifting the dark-gray zone (i.e., energy of the 2004 December 27 giant flare)
in Fig.~\ref{E_gf_E_max} to the left;
Moreover, the value of the power-law index $\alpha$ may be different
from $5/3$ (e.g., G{\"o}tz et al. 2006; $\alpha=1.9$), and the
luminosity function model is an approximation.
In reality the energy distribution
(i.e., $d{N}/d{E}\propto E^{-5/3}$) is probably not terminated abruptly
at $E_{max}$, as different SGRs may be born with different magnetic fields.
I note that although the tentative detection of a single
extragalactic SGR has a large Poisson uncertainty
($1_{-0.98}^{+4.6}$, at the $95\%$ CL),
the actual rate cannot be much smaller than one, or
it will not be consistent with the observed Galactic rate
of giant flares with energy above $3.7\times10^{46}$~erg.

Interestingly, the Galactic rate of giant flares
with energy above $2\times10^{44}$~erg
is $(40_{-20,-26}^{+47,+154})^{-1}$~yr$^{-1}$ per SGR
(The errors are the 1 and 2-$\sigma$ CL, respectively;
assuming three\footnote{These are: 1979 March 5 (SGR~0526$-$66);
1998 August 27 (SGR~1900$+$14); and 2004 December 27 (SGR~1806$-$20).
Note that, SGR flares are correlated (see also \S\ref{Disc})
and therefore Poisson statistics can be used only if the flares were originated
from different objects -- as in this case.}
giant flares were observed in 30~yrs).
This rate is indicated by the error-bar in
the left-hand side of Fig.~\ref{E_gf_E_max}.
The solid line is the $1$-$\sigma$ CI, and the dashed line
is the $2$-$\sigma$ CI implied by the Galactic rate.
In order to reconcile the Galactic rate
with the extragalactic rate,
a gradual break (or steepening) in the flare energy distribution is required
at $\lesssim 3\times10^{46}$~erg (at the $95\%$ CL).

I note that if I adopt the value of $\alpha=1.9$ found
by G\"{o}tz et al. (2006),
then the inferred rate
of giant flares with energy above $4\times10^{46}$~erg
will be even lower (by about $10\%$) than
the rate based on $\alpha=5/3$.
Moreover, for $\alpha=1.9$
I find that $E_{max} \lesssim 3\times10^{47}$,
at the $95\%$~CL.

To conclude, this analysis shows that:
$E_{max}\ltorder3\times10^{46}$~erg at the $95\%$ CL;
the rate of SGR giant flares is $(0.4-5)\times10^{-4}$~yr$^{-1}$ per SGR;
and the fraction of SGR flares among short-duration GRBs
cannot be much smaller than one percent,
otherwise it will be inconsistent with the Galactic rate.

\section{Discussion}
\label{Disc}

The fraction of SGR flares
among short GRBs and the rate of SGR giant flares
are closely related issues.
Measurements of these quantities are important for the understanding
of the SGR flare mechanism
and magnetic field strength.
Below I discuss
the possibility of observing additional extragalactic flares
and the implications of the measured SGR giant flare rate
for the magnetic field strength in the context of
the magnetar model.

The derived fraction of SGRs among short GRBs
implies that
about five to $\sim65$ extragalactic SGR flares were observed by BATSE.
Moreover, I estimate that the Swift satellite may find an
extragalactic SGR every several years,
and that the Gamma-Ray Large Area Space Telescope
will localize about $0.2$--$3$ extragalactic SGRs per year.
Based on the analysis presented in this paper,
it would be surprising if both 
GRB~000420B and
the recently discovered GRB~051103
(Golenetskii et al. 2005; Ofek et al. 2006; Frederiks et al. 2006)
were not associated with M74 and M81, respectively.

Verifying that a GRB is indeed an extragalactic SGR flare
requires either the detection of persistent X-ray emission,
which is beyond our current observational capabilities,
or the detection of additional flare activity.
The fact that SGR flares are correlated
may increase the probability of observing additional
flare activity from extragalactic SGRs.
The observed time between
low-energy SGR flares follows
a log-normal distribution
(Laros et al. 1987; Hurley et al. 1994; Cheng et al. 1996).
Therefore, it is more appropriate to
express the typical time between bursts using
the log-normal-mean, $M$, and -standard deviation, $S$.
Assuming a flare rate of $2\times10^{-4}$~yr$^{-1}$,
then $\tau$, the expectation value
of the time between giant flares,
with energy above $3.7\times10^{46}$~erg, is $5000$~yr.
Since the expectancy value of log-normal distribution
is given by 
$\exp(M+S^2/2)$, and assuming $S=3.46$ (Hurley et al. 1994)
is independent of energy,
the log mean for giant flares
is $M_{gf}\cong2.53$.
Interestingly, the cumulative log-normal distribution
suggests that given these parameters (i.e., $M_{gf}=2.53$; $S=3.46$),
the probability to observe a second giant flare of an SGR
within ten years from the first flare (given a sufficient energy source)
is  $47\%$.
Therefore, if
GRB~000420B is associated with M74,
and/or GRB~051103 is associated with M81, 
there is a fair chance to see another flare from these
galaxies in the next decade.
Furthermore, I note that 
if more than one
burst per SGR will be observed,
the use of Poisson statistics will
considerably overestimate the flare rate.

As discussed in \S\ref{Introduction},
Stella et al. (2005) argued that
about 70 giant flares,
with energy above $5\times10^{46}$~erg,
are expected 
from an SGR during its life-time,
and claimed that SGR
magnetic fields are in excess of $10^{16}$~G.
My analysis, however,
shows that the giant flare rate is two orders of magnitude smaller
than that estimated by Stella et al. (2005),
and therefore, magnetic fields of $B\approx10^{15}$~G
can provide the required energy source.
The giant flare rate I derive, 
$(0.4-5)\times10^{-4}$~yr$^{-1}$ per SGR,
suggests that
within the lifetime of an SGR, $10^{3}$-$10^{5}$~yrs,
it would have of order unity giant flares
with energy above $4\times10^{46}$~erg.

To summarize, I searched for a spatial coincidence of IPN GRBs
with galaxies within 20~Mpc.
I have found a single SGR candidate, possibly located within M74.
However, this match could be 
a chance coincidence.
Currently, the best way to test the hypothesis
that GRB~000420B arose in M74
is to detect additional SGR flares from this galaxy.
As SGR activity is correlated,
the expected waiting time to a second flare from the same SGR
is not necessarily long.
I use the results to place an upper limit on
the fraction of SGRs 
among short GRBs with fluence above
$\sim10^{-5}$~erg~cm$^{-2}$, of $<16\%$ at the $95\%$ CL.
This limit is consistent with the recent finding of
Nakar et al. (2006) and Lazzati et al. (2005),
but contrary to the latter paper it does not depend
on the assumption that all SGR giant flares
have the same light curve and spectrum.
I note that this limit is based on the assumption
that magnetars form predominantly in young stellar
populations (see however Levan et al. 2006).
I roughly estimate the efficiency of the IPN
to detect giant flares in each of the galaxies
in the sample, and
find
that the rate of giant flares,
with energy above $3.7\times10^{46}$~erg,
is $(0.4-5)\times10^{-4}$~yr$^{-1}$ per SGR.
Finally, I show that the SGR flare luminosity function is
consistent with a single power-law 
with a gradual cutoff or steepening
below $3\times10^{46}$~erg ($3\times10^{47}$~erg),
at the $95\%$ CL, for $\alpha=5/3$ ($1.9$).

\acknowledgments

I am grateful to Kevin Hurley for his assistance in my use of the
IPN data and for maintaining this important database,
and to an anonymous referee for useful comments.
I thank Shri Kulkarni, Ehud Nakar, Re'em Sari, Orly Gnat, 
Avishay Gal-Yam, and Brad Cenko, for valuable discussions.
This work is supported in part by grants from NSF and NASA.

\end{document}